\documentclass[12pt,preprint]{aastex63}

\newcommand\src{180916.J0158+65 }

\begin{document}

\title{Detection of Repeating FRB \src Down to Frequencies of 300 MHz}

\author[0000-0002-3426-7606]{P.~Chawla}
  \affiliation{Department of Physics, McGill University, 3600 rue University, Montr\'eal, QC H3A 2T8, Canada}
  \affiliation{McGill Space Institute, McGill University, 3550 rue University, Montr\'eal, QC H3A 2A7, Canada}
\author[0000-0001-5908-3152]{B.~C.~Andersen}
  \affiliation{Department of Physics, McGill University, 3600 rue University, Montr\'eal, QC H3A 2T8, Canada}
  \affiliation{McGill Space Institute, McGill University, 3550 rue University, Montr\'eal, QC H3A 2A7, Canada}
\author[0000-0002-3615-3514]{M.~Bhardwaj}
  \affiliation{Department of Physics, McGill University, 3600 rue University, Montr\'eal, QC H3A 2T8, Canada}
  \affiliation{McGill Space Institute, McGill University, 3550 rue University, Montr\'eal, QC H3A 2A7, Canada}
\author[0000-0001-8384-5049]{E.~Fonseca}
  \affiliation{Department of Physics, McGill University, 3600 rue University, Montr\'eal, QC H3A 2T8, Canada}
  \affiliation{McGill Space Institute, McGill University, 3550 rue University, Montr\'eal, QC H3A 2A7, Canada}
\author[0000-0003-3059-6223]{A.~Josephy}
  \affiliation{Department of Physics, McGill University, 3600 rue University, Montr\'eal, QC H3A 2T8, Canada}
  \affiliation{McGill Space Institute, McGill University, 3550 rue University, Montr\'eal, QC H3A 2A7, Canada}
\author[0000-0001-9345-0307]{V.~M.~Kaspi}
  \affiliation{Department of Physics, McGill University, 3600 rue University, Montr\'eal, QC H3A 2T8, Canada}
  \affiliation{McGill Space Institute, McGill University, 3550 rue University, Montr\'eal, QC H3A 2A7, Canada}
\author[0000-0002-2551-7554]{D.~Michilli}
  \affiliation{Department of Physics, McGill University, 3600 rue University, Montr\'eal, QC H3A 2T8, Canada}
  \affiliation{McGill Space Institute, McGill University, 3550 rue University, Montr\'eal, QC H3A 2A7, Canada}
\author[0000-0002-4795-697X]{Z.~Pleunis}
  \affiliation{Department of Physics, McGill University, 3600 rue University, Montr\'eal, QC H3A 2T8, Canada}
  \affiliation{McGill Space Institute, McGill University, 3550 rue University, Montr\'eal, QC H3A 2A7, Canada}
\author[0000-0003-3772-2798]{K.~M.~Bandura}
  \affiliation{CSEE, West Virginia University, Morgantown, WV 26505, USA}
  \affiliation{Center for Gravitational Waves and Cosmology, West Virginia University, Morgantown, WV 26505, USA}
\author[0000-0002-1429-9010]{C.~G.~Bassa}
  \affiliation{ASTRON, Netherlands Institute for Radio Astronomy, Oude Hoogeveensedijk 4, 7991 PD Dwingeloo, The Netherlands}
\author[0000-0001-8537-9299]{P.~J.~Boyle}
  \affiliation{Department of Physics, McGill University, 3600 rue University, Montr\'eal, QC H3A 2T8, Canada}
  \affiliation{McGill Space Institute, McGill University, 3550 rue University, Montr\'eal, QC H3A 2A7, Canada}
\author[0000-0002-1800-8233]{C.~Brar}
  \affiliation{Department of Physics, McGill University, 3600 rue University, Montr\'eal, QC H3A 2T8, Canada}
  \affiliation{McGill Space Institute, McGill University, 3550 rue University, Montr\'eal, QC H3A 2A7, Canada}  
\author[0000-0003-2047-5276]{T.~Cassanelli}
  \affiliation{Dunlap Institute for Astronomy \& Astrophysics, University of Toronto, 50 St.~George Street, Toronto, ON M5S 3H4, Canada}
  \affiliation{David A.~Dunlap Institute Department of Astronomy \& Astrophysics, University of Toronto, 50 St.~George Street, Toronto, ON M5S 3H4, Canada}
\author[0000-0003-2319-9676]{D.~Cubranic}
  \affiliation{Department of Physics \& Astronomy, University of British Columbia, 6224 Agricultural Road, Vancouver, BC V6T 1Z1, Canada}
\author{M.~Dobbs}
  \affiliation{Department of Physics, McGill University, 3600 rue University, Montr\'eal, QC H3A 2T8, Canada}
  \affiliation{McGill Space Institute, McGill University, 3550 rue University, Montr\'eal, QC H3A 2A7, Canada}
\author{F.~Q.~Dong}
  \affiliation{Department of Physics \& Astronomy, University of British Columbia, 6224 Agricultural Road, Vancouver, BC V6T 1Z1, Canada}
\author[0000-0002-3382-9558]{B.~M.~Gaensler}
  \affiliation{Dunlap Institute for Astronomy \& Astrophysics, University of Toronto, 50 St.~George Street, Toronto, ON M5S 3H4, Canada}
  \affiliation{David A.~Dunlap Institute Department of Astronomy \& Astrophysics, University of Toronto, 50 St.~George Street, Toronto, ON M5S 3H4, Canada}
\author[0000-0003-1884-348X]{D.~C.~Good}
  \affiliation{Department of Physics \& Astronomy, University of British Columbia, 6224 Agricultural Road, Vancouver, BC V6T 1Z1, Canada}
\author[0000-0003-2317-1446]{J.~W.~T.~Hessels}
  \affiliation{Anton Pannekoek Institute for Astronomy, University of Amsterdam, Science Park 904, 1098 XH, Amsterdam, The Netherlands}
  \affiliation{ASTRON, Netherlands Institute for Radio Astronomy, Oude Hoogeveensedijk 4, 7991 PD Dwingeloo, The Netherlands}
\author{T.~L.~Landecker}
    \affiliation{Dominion Radio Astrophysical Observatory, Herzberg Astronomy and Astrophysics Research Centre, National Research Council Canada, P.O. Box 248, Penticton, BC V2A 6J9, Canada}
\author[0000-0002-4209-7408]{C.~Leung}
  \affiliation{MIT Kavli Institute for Astrophysics and Space Research, Massachusetts Institute of Technology, 77 Massachusetts Ave, Cambridge, MA 02139, USA}
  \affiliation{Department of Physics, Massachusetts Institute of Technology, 77 Massachusetts Ave, Cambridge, MA 02139, USA}
\author[0000-0001-7931-0607]{D.~Z.~Li}
  \affiliation{Department of Physics, University of Toronto, 60 St.~George Street, Toronto, ON M5S 1A7, Canada}
  \affiliation{Canadian Institute for Theoretical Astrophysics, 60 St.~George Street, Toronto, ON M5S 3H8, Canada}
\author[0000-0001-7453-4273]{H.~-.~H.~Lin}
  \affiliation{Canadian Institute for Theoretical Astrophysics, 60 St.~George Street, Toronto, ON M5S 3H8, Canada}
  \affiliation{Max Planck Institute for Radio Astronomy, Auf dem Huegel 69, 53121 Bonn, Germany}
\author[0000-0002-4279-6946]{K.~Masui}
  \affiliation{MIT Kavli Institute for Astrophysics and Space Research, Massachusetts Institute of Technology, 77 Massachusetts Ave, Cambridge, MA 02139, USA}
  \affiliation{Department of Physics, Massachusetts Institute of Technology, 77 Massachusetts Ave, Cambridge, MA 02139, USA}
\author[0000-0001-7348-6900]{R.~Mckinven}
  \affiliation{Dunlap Institute for Astronomy \& Astrophysics, University of Toronto, 50 St.~George Street, Toronto, ON M5S 3H4, Canada}
  \affiliation{David A.~Dunlap Institute Department of Astronomy \& Astrophysics, University of Toronto, 50 St.~George Street, Toronto, ON M5S 3H4, Canada}
\author[0000-0002-0772-9326]{J.~Mena-Parra}
  \affiliation{MIT Kavli Institute for Astrophysics and Space Research, Massachusetts Institute of Technology, 77 Massachusetts Ave, Cambridge, MA 02139, USA}
\author[0000-0003-2095-0380]{M.~Merryfield}
  \affiliation{Department of Physics, McGill University, 3600 rue University, Montr\'eal, QC H3A 2T8, Canada}
  \affiliation{McGill Space Institute, McGill University, 3550 rue University, Montr\'eal, QC H3A 2A7, Canada}
\author[0000-0001-8845-1225]{B.~W.~Meyers}
  \affiliation{Department of Physics \& Astronomy, University of British Columbia, 6224 Agricultural Road, Vancouver, BC V6T 1Z1, Canada}
\author[0000-0002-9225-9428]{A.~Naidu}
  \affiliation{Department of Physics, McGill University, 3600 rue University, Montr\'eal, QC H3A 2T8, Canada}
  \affiliation{McGill Space Institute, McGill University, 3550 rue University, Montr\'eal, QC H3A 2A7, Canada}
\author[0000-0002-3616-5160]{C.~Ng}
   \affiliation{Dunlap Institute for Astronomy \& Astrophysics, University of Toronto, 50 St.~George Street, Toronto, ON M5S 3H4, Canada}
\author{C.~Patel}
  \affiliation{Dunlap Institute for Astronomy \& Astrophysics, University of Toronto, 50 St.~George Street, Toronto, ON M5S 3H4, Canada}
  \affiliation{Department of Physics, McGill University, 3600 rue University, Montr\'eal, QC H3A 2T8, Canada}
\author{M.~Rafiei-Ravandi}
  \affiliation{Perimeter Institute for Theoretical Physics, 31 Caroline Street N, Waterloo, ON N2L 2Y5, Canada}
\author[0000-0003-1842-6096]{M.~Rahman}
  \affiliation{Dunlap Institute for Astronomy \& Astrophysics, University of Toronto, 50 St.~George Street, Toronto, ON M5S 3H4, Canada}
\author[0000-0001-5504-229X]{P.~Sanghavi}
  \affiliation{CSEE, West Virginia University, Morgantown, WV 26505, USA}
  \affiliation{Center for Gravitational Waves and Cosmology, West Virginia University, Morgantown, WV 26505, USA}
\author[0000-0002-7374-7119]{P.~Scholz}
  \affiliation{Dunlap Institute for Astronomy \& Astrophysics, University of Toronto, 50 St.~George Street, Toronto, ON M5S 3H4, Canada}
\author[0000-0002-4209-7408]{K.~Shin}
  \affiliation{MIT Kavli Institute for Astrophysics and Space Research, Massachusetts Institute of Technology, 77 Massachusetts Ave, Cambridge, MA 02139, USA}
  \affiliation{Department of Physics, Massachusetts Institute of Technology, 77 Massachusetts Ave, Cambridge, MA 02139, USA}  
\author{K.~M.~Smith}
  \affiliation{Perimeter Institute for Theoretical Physics, 31 Caroline Street N, Waterloo, ON N2L 2Y5, Canada}
\author[0000-0001-9784-8670]{I.~H.~Stairs}
  \affiliation{Department of Physics \& Astronomy, University of British Columbia, 6224 Agricultural Road, Vancouver, BC V6T 1Z1, Canada}
\author[0000-0003-2548-2926]{S.~P.~Tendulkar}
  \affiliation{Department of Physics, McGill University, 3600 rue University, Montr\'eal, QC H3A 2T8, Canada}
  \affiliation{McGill Space Institute, McGill University, 3550 rue University, Montr\'eal, QC H3A 2A7, Canada}
\author[0000-0003-4535-9378]{K.~Vanderlinde}
  \affiliation{Dunlap Institute for Astronomy \& Astrophysics, University of Toronto, 50 St.~George Street, Toronto, ON M5S 3H4, Canada}
  \affiliation{David A.~Dunlap Institute Department of Astronomy \& Astrophysics, University of Toronto, 50 St.~George Street, Toronto, ON M5S 3H4, Canada}
  
\correspondingauthor{P. Chawla}
\email{pragya.chawla@mail.mcgill.ca}

\begin{abstract}
We report on the detection of seven bursts from the periodically active,
repeating fast radio burst (FRB) source FRB \src in the 300--400-MHz frequency range with the Green Bank Telescope (GBT). 
Emission in multiple bursts is visible down to the bottom of the GBT band, suggesting that the cutoff frequency (if it exists) for FRB emission
is
lower than
300 MHz. Observations were conducted during predicted periods of activity of the source, and had simultaneous coverage with the Low Frequency Array (LOFAR) and the FRB backend on the Canadian Hydrogen Intensity Mapping Experiment (CHIME) telescope. We find that one of the GBT-detected bursts has potentially associated emission in the CHIME band (400--800 MHz) but we detect no bursts in the LOFAR band (110--190 MHz), placing a limit of $\alpha > -1.0$ on the spectral index of broadband emission from the source.
We also find that emission from the source is severely band-limited with burst bandwidths as low as $\sim$40 MHz.
In addition, we place the strictest constraint on observable scattering of the source, $<$ 1.7 ms, at 350 MHz, suggesting that the circumburst environment does not have strong scattering properties. Additionally, knowing that the circumburst environment is optically thin to free-free absorption at 300 MHz, we find evidence against the association of a hyper-compact H\textsc{ii} region or a young supernova remnant (age $<$ 50 yr) with the source.
\end{abstract}

\section{Introduction} \label{sec:intro}
Fast radio bursts (FRBs) are bright,
millisecond-duration radio
transients
of unknown physical origin 
(see \citealt{petroff2019} and \citealt{cordes2019} for reviews). Their measured dispersion measures (DMs) 
are in excess of those expected
from the Milky Way, suggesting that FRBs are located at cosmological distances.
Identification of the host galaxy
for five of the 110 published FRBs\footnote{\url{http://frbcat.org}} \citep{petroff2016} has confirmed their cosmological origin, placing these five sources at redshifts between 0.03 and 0.66 \citep{tendulkar2017,bannister2019,ravi2019,prochaska2019,marcote2020}. 
Dozens of emission models involving progenitors ranging from compact objects to cosmic strings have been proposed to explain the inferred
isotropic energy output of $\sim$10$^{40}$ erg for these sources
(see \citealt{platts2019} for a summary of the proposed models\footnote{\url{http://frbtheorycat.org}}). 

Over the past decade, most FRBs have been observed at a frequency of $\sim$1 GHz, with one of the sources, FRB 121102, also being detected 
in the 4--8 GHz frequency range \citep{michilli2018,gajjar2018}. However, the recent detection of FRBs by the Canadian Hydrogen Intensity Mapping Experiment telescope (CHIME) in the frequency range of 400--800 MHz \citep{chime2019a} has shown that FRB emission 
extends to lower radio frequencies. With the apparent ubiquity of these bursts above a frequency of 400 MHz, it is surprising that none of the FRB searches conducted in the frequency range of 300--400 MHz (e.g., \citealt{deneva2016}, \citealt{chawla2017}, \citealt{rajwade2020limits}) have detected any FRBs so far.  

Observations of FRBs at low frequencies are important to ascertain whether there exists a
cutoff or turnover
frequency for their emission, which can help constrain proposed emission mechanisms. FRBs could also be rendered undetectable at low frequencies due to a spectral turnover arising from propagation effects in the circumburst environment
or due to scattering (multipath propagation of signals caused by an intervening ionized medium). 
Either 
of these possibilities offers an opportunity to constrain the properties of the circumburst medium using low-frequency detections \citep{chime2019a,ravi2019}. Additionally, observations in the 300--400-MHz frequency range can enable studies of the frequency dependence of FRB activity, allowing determination of optimal observing strategies for future FRB searches with instruments such as the Low-Frequency Array (LOFAR) and the Murchison Widefield Array (MWA) which have not detected any FRBs so far at frequencies close to 150 MHz \citep{coenen2014,tingay2015,karastergiou2015,sokolowski2018no}.

To detect emission from FRBs at low frequencies, a complementary strategy to blind searches is targeted follow-up of repeating FRBs (e.g., \citealt{houben2019}). There are currently 20 known repeating FRBs \citep{spitler2016,chime2019b,chime2019c,kumar2019,fonseca2020}. It is as yet unclear whether these sources have a different physical origin as compared to the so-far non-repeating FRBs. Burst widths for repeating sources are found to be larger, on average, than 
those for the so-far non-repeating sources.
While this observation supports the notion of potentially different emission mechanisms and/or local environments for the two \citep{scholz2016,chime2019c,fonseca2020}, 
analysis of volumetric occurrence rates suggests that a large fraction of FRBs must repeat \citep{ravi2019b}.

The CHIME/FRB collaboration has recently reported on the detection of a 16.35-day modulation (or alias thereof) in the activity of one of the repeating FRB sources, FRB \src \citep{chime2020} with bursts from this source
being detected in a 5-day-long activity window.
This source is also interesting because of its recent localization to a star-forming region in a massive spiral galaxy at a redshift $z = 0.0337 \pm 0.0002$ (corresponding to a luminosity distance of 149.0 $\pm$ 0.9 Mpc), making it the closest of any of the localized FRBs \citep{marcote2020}. 

Here we report on the detection of low-frequency radio emission from FRB \src with the 100-m diameter Robert C. Byrd Green Bank Telescope (GBT) and simultaneous observations of the source with the CHIME/FRB instrument and LOFAR. In Section \ref{sec:observations}, we provide details of observations made with the above-mentioned instruments. In Section \ref{sec:analysis}, we describe the methodology used to determine burst properties and sensitivity thresholds for these observations. In Section \ref{sec:discussion}, we discuss the frequency dependence of burst activity as well as the implications of the detection of emission at low frequencies from this source. We summarize our conclusions in Section \ref{sec:conclusion}.   

\section{Observations} \label{sec:observations}
\subsection{Green Bank Telescope}
\label{sec:GBT}
We observed the interferometric position of FRB \src (RA = 01$^\mathrm{h}$58$^\mathrm{m}$00$\fs$0075, Dec. = 65$^\circ$43$'$00.3152$''$; \citealt{marcote2020}), in the period from 2019 November 15 to 2020 January 20,
with the GBT at a central frequency of 350 MHz. Data spanning 100 MHz of bandwidth were recorded using the Green Bank Ultimate Pulsar Processing Instrument backend (GUPPI; \citealt{duplain2008}). We conducted six observations of the source, all of which were within a $\pm$1.2-day interval of predicted epochs of peak source activity \citep{chime2020}. Details of the observations are provided in Table \ref{tab:observations} with a timeline shown in Figure \ref{fig:exposure}.  

In each observation, data for all four Stokes parameters were recorded for 512 frequency channels at a cadence of 20.48 $\mu$s.  Since the DM of the source was known prior to the observations, data were coherently dedispersed to the nominal DM (349.5 pc cm$^{-3}$, \citealt{chime2019c}), thereby increasing search sensitivity by mitigating the effects of intra-channel dispersive smearing. 

We downsampled the data to a resolution of 327.68 $\mu$s and searched for bursts using the PRESTO software package\footnote{\url{https://github.com/scottransom/presto}} \citep{ransom2001}. In order to do so, we masked time samples and frequency channels containing radio frequency interference (RFI) using PRESTO's \texttt{rfifind}. This process reduced the usable bandwidth in each observation to $\sim$80 MHz due to persistent RFI in the 360--380 MHz frequency range at the telescope site. We then dedispersed the data at a large number of trial DMs (in steps of 0.03 pc cm$^{-3}$) in a narrow range of 20 pc cm$^{-3}$ around the nominal DM. Each dedispersed topocentric time series was searched for single pulses (with a maximum width of 100 ms) using the PRESTO-based matched filtering algorithm,  \texttt{single\_pulse\_search.py}. We found a total of eight candidate bursts having S/N $>7$ in these observations, seven of which were found to be astrophysical after examining their dynamic spectra. The other candidate burst comprised of two narrowband RFI signals which were observed at different times but were detected with a high S/N as they coincidentally lined up at the DM of the source. The dynamic spectra of these bursts are shown in Figure~\ref{fig:waterfalls} with the burst properties determined in Section \ref{sec:analysis}.

\subsection{CHIME/FRB Instrument}
\label{sec:CHIME}
Four of the six GBT observations were scheduled to coincide with the transit of the source over the CHIME telescope. CHIME is a transit telescope sensitive in the frequency range of 400--800 MHz with the CHIME/FRB instrument searching for dispersed signals in 1,024 synthesized beams using a real-time detection pipeline \citep{chime2018}. The repeating FRB \src is observable daily for 12 minutes within the FWHM (at 600 MHz) of the synthesized beams of the CHIME/FRB system. Transit of the source over the extent of the primary beam of the CHIME telescope is much longer ($\sim$40 minutes). However, sensitivity to the source varies significantly during the primary beam transit. 

\begin{figure}
    \centering
    \includegraphics[scale=0.8]{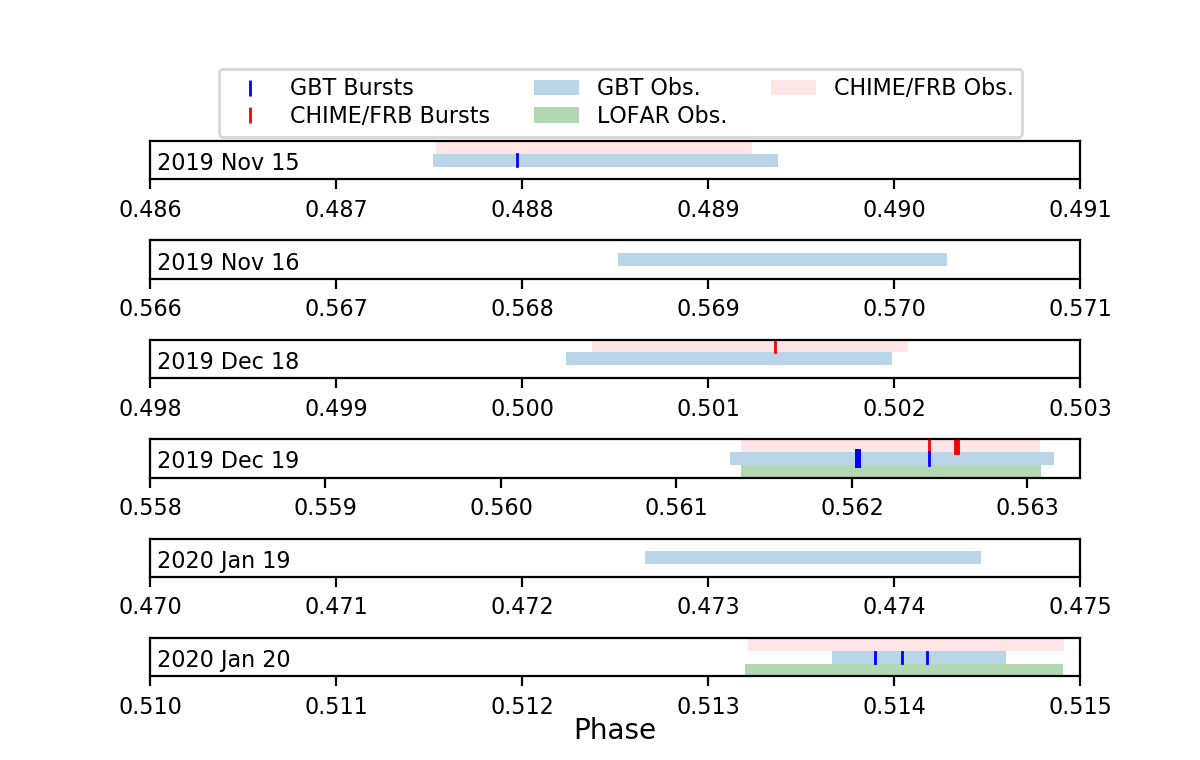}
    \caption{The exposure of GBT, LOFAR and the CHIME/FRB system for each day of observations as a function of source activity phase (see \S\ref{sec:phase}). The duration of each observation is represented by the shaded regions with the vertical lines marking the burst detection times. Heavy lines represent detection of two bursts.}
    \label{fig:exposure}
\end{figure}

All CHIME/FRB compute nodes responsible for processing data for the four synthesized beams through which the source transits
were operational at the time of the GBT observations. Four bursts from the source were detected by the CHIME/FRB system during these observations \citep{chime2020}, three of which do not have any coincident emission detected with the GBT (see Figure \ref{fig:exposure}). However, one of these bursts, with emission in a narrow frequency range of $\sim$50 MHz at the bottom of the CHIME band, is found to be associated with the GBT-detected burst 191219C after referencing burst arrival times for both telescopes to the solar system barycenter. We caution that we cannot accurately determine the temporal separation between the two bursts as there could be as-yet-uncharacterized offsets (of the order of $\sim$ms) between the timestamps reported by the two backends (GUPPI and CHIME/FRB). However, our current estimate of the separation between the best-fit burst arrival times (see Table \ref{tab:bursts}) is $\sim$23 ms, which implies that the bursts together constitute emission drifting downwards in frequency --- even after correction for dispersion. The phenomenon of downward-drifting sub-bursts has been observed previously for this source as well as for other repeating FRBs \citep{2019ApJ...876L..23H, chime2019b,chime2019c}. The dynamic spectra of these bursts are shown in Figure \ref{fig:waterfalls} with characterization of burst properties and the frequency-drift rate described in Section \ref{sec:analysis}.

\floattable
\begin{deluxetable}{ccccccc}
\tablenum{1}
\caption{Summary of Observations\label{tab:observations}}
\tablewidth{0pt}
\tablehead{
\colhead{Telescope} & \colhead{Obs. Freq.} & \colhead{Bandwidth} &
\colhead{Calendar Date} &
\colhead{Start Time} &
\colhead{Duration} &
\colhead{No. of} \\
\colhead{} & 
\colhead{(MHz)} &
\colhead{(MHz)} &
\colhead{}
& \colhead{(MJD)\tablenotemark{\textrm{a}}} 
& \colhead{(hours)} 
& \colhead{Detections\tablenotemark{\textrm{b}}} 
}
\startdata
LOFAR & 150 & 80 & 2019 Dec 19 & 58836.15841 & 0.67 & ...\\ 
 & & & 2020 Jan 20 & 58868.07083 & 0.67 & ...\\
GBT & 350 & 100 & 2019 Nov 15 & 58802.25100 & 0.73 & 1\\
 & & & 2019 Nov 16 & 58803.57524 & 0.69 & ...\\
 & & & 2019 Dec 18 & 58835.15885 & 0.69 & ...\\
 & & & 2019 Dec 19 & 58836.15731 & 0.73 & 3\\
 & & & 2020 Jan 19 & 58867.40803 & 0.71 & ...\\
 & & & 2020 Jan 20 & 58868.07846 & 0.37 & 3\\
CHIME/FRB\tablenotemark{\textrm{c}} & 600 & 400 & 2019 Nov 15 & 58802.25126 & 0.67 & ...\\
& & & 2019 Dec 18 & 58835.16114 & 0.67 & 1\\
& & & 2019 Dec 19 & 58836.15841 & 0.67 & 3 \\
& & & 2020 Jan 20 & 58868.07103 & 0.67 & ...\\
\enddata
\tablenotetext{a}{Reported start times are topocentric at each observing site.}
\tablenotetext{b}{Detections with the CHIME/FRB system have been reported on by \citet{chime2020}.}
\tablenotetext{c}{The start time and duration of each observation with the CHIME/FRB system corresponds to the transit of the source across the primary beam of the telescope (see \S\ref{sec:CHIME})}
\end{deluxetable}
   
\subsection{Low-Frequency Array (LOFAR)}
\label{sec:LOFAR}

We recorded beam-formed complex voltage data with the LOFAR High Band Antennas of all Core stations \citep{2013A&A...556A...2V, 2011A&A...530A..80S} at 110--190 MHz during the time of two GBT observations in which bursts are detected. This data has a native resolution of 5.12 $\mu$s and 0.195312 MHz.

All channels were coherently dedispersed to 350 pc cm$^{-3}$ using \texttt{cdmt} \citep{2017A&C....18...40B}, RFI was masked using PRESTO’s \texttt{rfifind} and the barycentered start times of the observations were calculated using \texttt{sigproc}’s \texttt{barycentre} module. 

We performed a traditional single-pulse search of the observations, using PRESTO’s \texttt{single\_pulse\_search.py}, split into a high-time resolution and a low-time resolution search. We adopt this strategy to be sensitive to both narrow  (similar burst width as in the GBT band) and wide bursts, in case bursts are intrinsically broader at lower frequencies or are broadened by scattering. For the high-time-resolution search, the data were downsampled to $t_s = 0.65536$ ms, and DMs 330--370~pc~cm$^{-3}$ were searched in 0.005~pc~cm$^{-3}$ steps up to burst widths of 0.05 s. For the low-time-resolution-search, the data were downsampled to $t_s = 20.97152$ ms, and DMs 330--370~pc~cm$^{-3}$ were searched in 0.2~pc~cm$^{-3}$ steps up to burst widths of 0.3~s. No bursts were detected with S/N $> 7$ and $> 9$, the noise floors for the high and low-time resolution searches, respectively.

We converted the GBT burst times-of-arrival (TOAs; listed in Table \ref{tab:bursts}) to expected TOAs at 150 MHz in the barycentric frame and generated $\sim$20 s of dynamic spectra around that time, incoherently dedispersed to 348.82 pc cm$^{-3}$ \citep[the best measured DM from high-resolution CHIME/FRB data;][]{chime2020}, sub-banded to 200 channels of 0.390624 MHz bandwidth each and downsampled to 1.96608-ms and 20.97152-ms time resolution. We inspected these dynamic spectra by eye in order to be sensitive to bursts with potentially narrow ($\sim$10--20 MHz) emission bandwidths and slow frequency drifts, to which a traditional search that sums over the full bandwidth has reduced sensitivity. This burst spectrum
is expected from extrapolating burst widths and linear drift rates at higher frequencies down to 150 MHz \citep{2019ApJ...876L..23H, josephy2019}. No bursts were found.

A 300-s test observation of pulsar B2111+46 (with a spin period of 1.01 s) was recorded and processed in the way described above (except that it was dedispersed at both the coherent and incoherent dedispersion stage to 141.3 pc cm$^{-3}$). Pulses were detected in both a by-eye inspection of dynamic spectra and in a blind search, as expected. As the closest test observation is from 2019 July 30 this gives confidence in the processing pipeline.

\section{Analysis} \label{sec:analysis}
\subsection{Determination of Burst Properties}\label{sec:properties}
\begin{figure}
    \centering
    \includegraphics[scale=0.7]{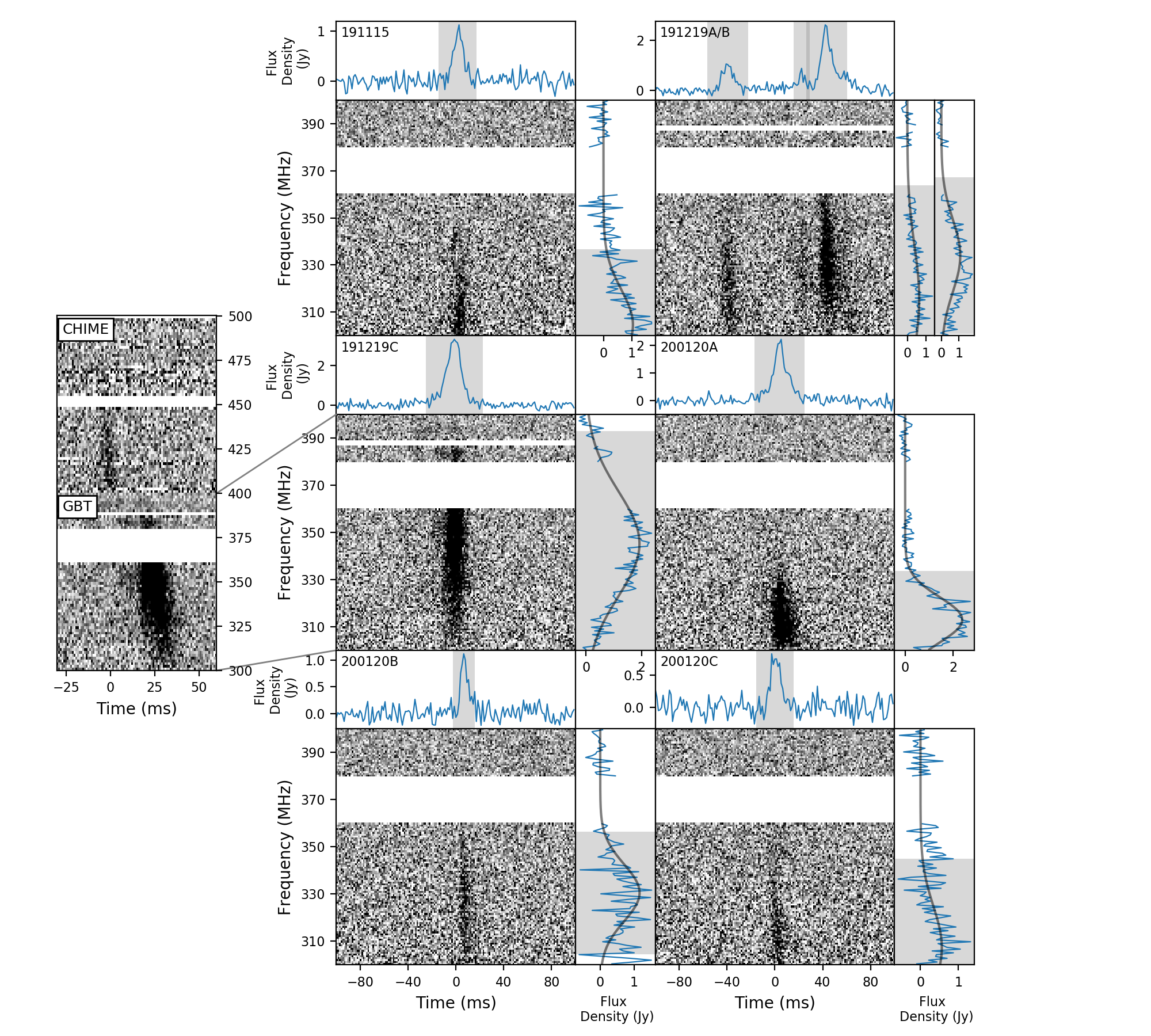}
    \vspace{-15pt}
    \caption{Dynamic spectrum (``waterfall") plots of seven bursts from the repeating FRB \src detected with the GBT. For each burst, data for 128 frequency sub-bands, corrected for the receiver bandpass and dedispersed to the structure-optimizing DM (listed in Table \ref{tab:bursts}), are plotted here. In order to better visualize the burst, each dynamic spectrum is downsampled to a resolution of 1.31 ms and intensity values are saturated at the 5th and 95th percentiles. Horizontal white bands are the frequency channels masked due to RFI. The band-averaged time series is plotted on the top of each dynamic spectrum. The on-pulse spectrum (plotted on the right) is obtained by averaging the flux within twice the FWTM of a Gaussian model fit to the time series. The assumed on-pulse region is shaded in grey in the top panel with the shaded region in the right panel showing the FWTM of a Gaussian model fit to the on-pulse spectrum.  Dynamic spectra for bursts 191219A and 191219B are shown in the same panel since the arrival times for the two are within $\sim$60 ms of each other. {\it Burst 191219C sub-panel}:
    This burst was associated with a CHIME/FRB detection (see \S\ref{sec:CHIME}). The composite dynamic spectrum for the two bursts, shown in the left-most panel, is corrected for the instrument bandpass and has a time and frequency resolution of 0.98304 ms and 1.5625 MHz, respectively.
    Data for the composite spectrum are dedispersed to the average DM of the source, 348.82 pc cm$^{-3}$, whereas data shown in the panel to the right are dedispersed to the structure-optimizing DM derived from the GBT-detection (349.5 pc cm$^{-3}$; see \S\ref{sec:properties}). There is no visible emission in the CHIME/FRB data outside the frequency range plotted here.}
    \label{fig:waterfalls}
\end{figure}

We determine burst properties from data which are downsampled to a resolution of 327.68 $\mu$s
and corrected for the receiver bandpass using the radiometer equation 
(see, e.g., \citealt{lorimer2005}), 
\begin{equation}\label{eq:radiometer}
    \Delta S_\textrm{sys} = \frac{T_\textrm{rec}+T_\textrm{sky}}{G \sqrt{n_\textrm{p} t_\textrm{s} \Delta f}}, 
\end{equation}
where $\Delta S_\textrm{sys}$ is the rms noise for each frequency channel, $T_\textrm{rec}$ is the receiver temperature, 
$G$ is the telescope gain, $n_\textrm{p}$ is the number of summed polarizations, $t_\textrm{s}$ is the sampling time and $\Delta f$ is the frequency resolution. The GBT's 350-MHz receiver has\footnote{\url{https://science.nrao.edu/facilities/gbt/proposing/GBTpg.pdf}} $n_\textrm{p} = 2$, $T_\textrm{rec} = 23$ K and $G = $ 2 K Jy$^{-1}$. The sky temperature at the central frequency of each channel, $T_\textrm{sky}$, was estimated using the 408-MHz all-sky map produced by \citet{remazeilles2015} by assuming a spectral index of $-$2.55 for Galactic synchrotron emission \citep{haslam1982}. We calibrate the data by subtracting the off-pulse mean, dividing by the off-pulse standard deviation and converting the counts in each frequency channel to a flux density
using Equation \ref{eq:radiometer}. 
Here we define the on-pulse region (shown in Figure \ref{fig:waterfalls}) as twice the full width at tenth maximum (FWTM) in a Gaussian model fit to the band-averaged time series. 

Using the calibrated data, we estimate burst DMs with the
\texttt{DM\_phase} package\footnote{\url{https://github.com/danielemichilli/DM_phase}} \citep{seymour2019} 
which has previously been used to characterize 
FRBs with complex morphologies, particularly those with downward-drifting sub-bursts. 
For each burst, the algorithm estimates the optimal DM based on whether the coherent power over the emission bandwidth, 
and hence the burst structure, is maximized. 
The measured DMs (see Table~\ref{tab:bursts}) for all but two GBT detections are consistent within 1$\sigma$ uncertainties 
with the average DM of 348.82 pc cm$^{-3}$ measured for the source using complex voltage data recorded by the CHIME/FRB system \citep{chime2020}. 
The DMs of the other two detections, 191115 and 191219C, although higher, are
consistent with the average source DM at the 3$\sigma$ level. 
We note that measurements can be biased high
if drifting sub-structure within bursts is unresolved at the time resolution 
used for obtaining the best-fit DM.  

We also fitted models of dynamic spectra to all calibrated GBT bursts for estimating component widths and 
scattering timescales, using the same least-squares algorithm employed in previous CHIME/FRB analyses \citep{chime2019a,chime2019b,chime2019c,fonseca2020}. For all modeling, we held the DM fixed to the values listed in Table~\ref{tab:bursts} while fitting six parameters that describe a pulse broadening function -- the convolution of a Gaussian temporal profile with a one-sided exponential decay induced by scattering \citep[e.g.][]{mckinnon2014} --- modulated by a running power law across the band. However, FRBs from repeating sources regularly display complex behavior, such as 
drifting sub-structure 
as well as frequency-dependent profile variations intrinsic to the emission mechanism \citep{2019ApJ...876L..23H}. In the low-S/N limit, such frequency-dependent structure can be temporally unresolved and mimic the effects of scatter broadening, potentially biasing the 
scattering measurements. However, in order to assess the statistical significance of scattering and place upper limits whenever possible, we nonetheless apply two-dimensional pulse broadening functions as models to the GBT spectra in a manner consistent with previous CHIME/FRB analyses. 

We estimate fluence by integrating the extent of each burst in the band-averaged time series 
with the peak flux density estimated to be the highest value in this time series. Both these measurements rely
on the off-pulse mean and standard deviation used in correcting the data for the receiver bandpass. Therefore, we generate different realizations of calibrated data, each using the off-pulse mean and standard deviation from a different time chunk (of 3-s duration) in the observation in which the burst was detected. We measure the fluence and peak flux density for each realization of the calibrated data and report the mean and standard deviation of these measurements in Table \ref{tab:bursts}. 

\floattable
\begin{deluxetable}{ccccccccccc}
\rotate
\tablenum{2}
\caption{Burst Properties\tablenotemark{\textrm{a}}\label{tab:bursts}}
\tablehead{
\colhead{Name} & \colhead{Arrival Time\tablenotemark{\textrm{b}}} & 
\colhead{Detection} &
\colhead{DM\tablenotemark{\textrm{c}}} &
\colhead{Scattering} &
\colhead{Temporal Width\tablenotemark{\textrm{e}}} &
\colhead{Fluence} &
\colhead{Peak Flux} & 
\colhead{Center} &
\colhead{Bandwidth} &
\colhead{RM\tablenotemark{\textrm{g}}}\\
\colhead{} & 
\colhead{(MJD)} &
\colhead{S/N} &
\colhead{(pc cm$^{-3}$)} 
& \colhead{Time (ms)\tablenotemark{\textrm{d}}} 
& \colhead{(ms)} 
& \colhead{(Jy ms)} 
& \colhead{Density\tablenotemark{\textrm{f}}}
& \colhead{Freq.}
& \colhead{(FWTM; MHz)}
& \colhead{(rad m$^{-2}$)}\\
\colhead{} & 
\colhead{} &
\colhead{} &
\colhead{} 
& \colhead{} 
& \colhead{} 
& \colhead{} 
& \colhead{(Jy)}
& \colhead{(MHz)}
& \colhead{}
& \colhead{}}
\startdata
191115 & 58802.25840267 & 11.3 & 349.3(2) & $<$2.9 & 3.6(5) & 10.2(5) & 1.18(6) & 304(5) & 64(17) & \\
191219A & 58836.16929624 & 13.3 & 348.8(4) & 5.9(3) & 2.7(4) & 11.2(4) & 1.24(5) & 316(5) & 96(19) &  $-$116.9(5) \\
191219B\tablenotemark{\textrm{h}} & 58836.16929695 & 28.6 & 348.8(4) & 5.9(3) & 3.9(8) & 5.2(2) & 1.09(4) & 334(1) & 66(6) & $-$116.9(2) \\
& 58836.16929720 & & & & 2.42(18) & 31.9(1.1) & 2.66(9) & & & \\
191219C & 58836.17591822 & 54.9 & 349.5(3) & $<$1.7 & 5.89(11) & 48.9(1.7) & 3.58(13) & 345.2(7) & 96(2) & $-$116.6(2) \\
200120A & 58868.08221442 & 19.3 & 348.9(1) & 4.1(3) & 3.7(2) & 28.1(9)	& 2.28(7) &	312.9(6) & 41(2) & $-$117.7(3) \\
200120B & 58868.08461892 & 14.4 & 348.7(2) & 3.1(5) & 1.5(3) & 7.8(2) & 1.30(4) & 330.5(9) & 52(4) & \\
200120C & 58868.08679636 & 11.9 & 348.8(2) & 1.8(9) & 3.2(6) & 7.7(2) & 0.98(3) & 307(5) & 75(16) & \\
\enddata
\tablenotetext{a}{Uncertainties and upper limits are reported at the $1\sigma$ and $2\sigma$ confidence level, respectively.}
\tablenotetext{b}{Arrival times are corrected to the solar system barycenter and referenced to infinite frequency.}
\tablenotetext{c}{DMs reported here are obtained through the process of structure-optimization (see \S\ref{sec:analysis}) with the dynamic spectra shown in Figure \ref{fig:waterfalls} dedispersed to these DMs.}
\tablenotetext{d}{Scattering timescales reported here are measured at a frequency of 350 MHz. See Section \ref{sec:scattering} for a discussion on the scattering estimates and sources of bias that can produce apparent variations.}
\tablenotetext{e}{All values are intrinsic burst widths since spectra are fitted with two-dimensional pulse broadening functions, and thus separate frequency-dependent scattering contributions and intrinsic components from the observed widths.}
\tablenotetext{f}{Peak flux density for each burst is estimated from the corresponding band-averaged time series at a resolution of 327.68~$\mu$s.}
\tablenotetext{g}{The values are not corrected for ionospheric contribution.}
\tablenotetext{h}{Burst 191219B is treated as having two components rather than being two separate bursts since the emission between the two does not revert to the baseline noise level. Therefore, arrival time, width, fluence and peak flux density for the two components are reported separately.}
\end{deluxetable}

\subsection{Drift Rate Measurement for Burst 191219C}
If the estimate of the temporal separation between burst 191219C and its coincident CHIME/FRB detection is correct, then we can measure the linear drift rate of the sub-bursts in the burst envelope that drifts from the bottom of the CHIME band into the top of the GBT band (inset panel on the left in Fig.~\ref{fig:waterfalls}) using a Monte Carlo resampled auto-correlation analysis \citep{chime2019b}. We use the composite dynamic spectrum in the frequency range of 300--500 MHz having 256 normalized frequency channels dedispersed to 348.82 pc cm$^{-3}$ for this analysis. We use 100 $\times$ 100 = 10,000 random noise and DM uncertainty samples, and set the difference between the best known DM for the source and the inverse-variance weighted average DM of the GBT bursts, 0.12 pc cm$^{-3}$, as the DM uncertainty. Two systematics in the auto-correlation analysis are the 0.1953125 MHz offset between the two bands and the alignment of time samples of the two systems, which is uncertain on the $\sim$30 $\mu$s level. Both effects introduce uncertainties on the drift rate measurement that are much smaller than the nominal uncertainty calculated from resampling the noise and DM uncertainty distributions. However, as noted above, there might be an additional unaccounted for offset between the two receivers on the $\sim$ms level. We measure d$\nu$/dt $= -4.2^{+0.4}_{-0.4}$ MHz ms$^{-1}$ (68\% confidence interval), under the assumption that that receiver time offset is negligible. This value is in line with the general observed trend for the drift rates to be lower at lower frequencies for bursts from FRB 121102 \citep{2019ApJ...876L..23H, josephy2019} and it points to a similar trend for FRB 180916.J0158+65. However, for FRB 180916.J0158+65, the lack of drift rate measurements at gigahertz
frequencies impedes a measurement of drift rate evolution with frequency for now.

\subsection{Polarization Properties}
\begin{figure*}
    \centering
    \includegraphics[width=\textwidth]{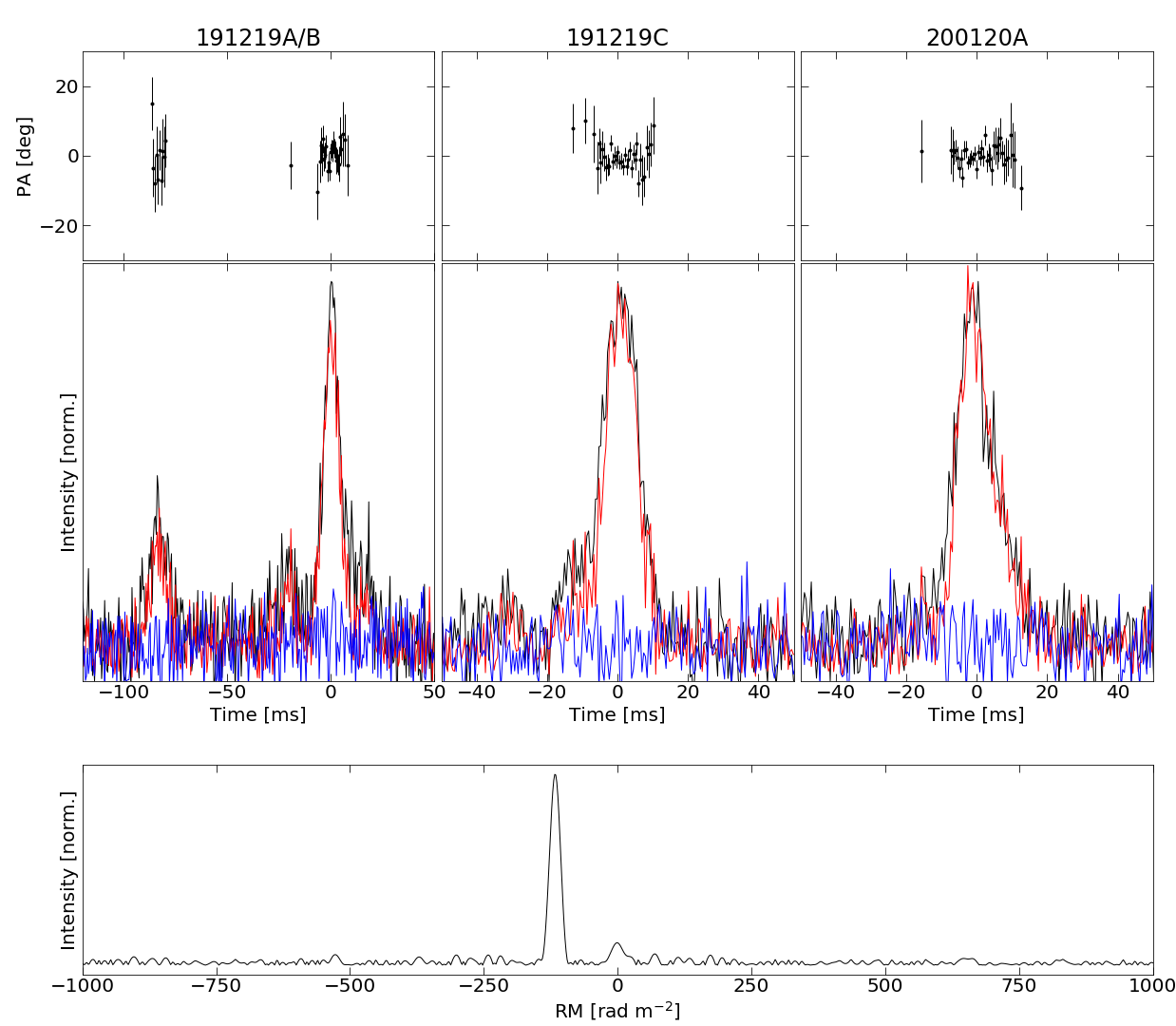}
    \caption{Polarization properties of four of the GBT bursts. Top: Pulse profiles, in normalized units, of total intensity (black), linear polarization (red) and circular polarization (blue) with polarization angles plotted on top. Polarization angle measurements have been rotated independently in each panel to have a null mean. Bottom: example of the FDF for burst 191219C after applying the cleaning algorithm (see text).}
    \label{fig:pol}
\end{figure*}

Polarization properties have been analyzed for four bursts of our sample. 
The lower S/N of the rest of the detections did not allow us to obtain robust measurements.
One second of frequency-resolved Stokes parameters have been extracted for each burst by using the \texttt{PSRCHIVE} library\footnote{\url{http://psrchive.sourceforge.net}} \citep{hotan2004}. 
A one-minute scan of a noise diode was acquired at the beginning of each observation and was used to calibrate the Stokes parameters of the bursts \citep{vanstraten2012}. 
All the bursts have been dedispersed to $349.5$\,pc\,cm$^{-3}$ for this analysis.
Following \citet{chime2019b}, the rotation measures (RMs) of the bursts have been calculated by using the implementation of rotation
measure synthesis \citep{burn1966,brentjens2005} in the RM-tools package\footnote{\url{https://github.com/CIRADA-Tools/RM}} and the resulting Faraday dispersion function (FDF) was cleaned with the deconvolution algorithm presented by \citet{heald2009}.
The FDF of bursts detected on MJD 58836 showed effects of a bad calibration, apparent in the presence of peaks at RM values symmetric about zero.
This was solved by using the scan of the noise diode acquired on MJD 58835.  The quality of the FDFs, one of which is reported in the bottom of Figure \ref{fig:pol}, demonstrates the validity of this operation.

The measured RM values are reported in Table~\ref{tab:bursts}.
On 2018 December 26, the RM for FRB 180916.J0158+65 was measured to be $-114.6(6)$\,rad\,m$^{-2}$ \citep{chime2019b}; it therefore shows possible evidence of a decrease of 2--3\,rad\,m$^{-2}$ over one year.
According to the RM model of the ionosphere \texttt{RMextract} \citep{2018ascl.soft06024M}, its varying contribution can account for $\lesssim0.4$\,rad\,m$^{-2}$.
However, systematic effects between GBT and CHIME could account for the remaining discrepancy.
Among the GBT bursts, 191219A and 191219B are separated by less than 100\,ms. They show the same RM within the errors. 
Burst 191219C, detected on the same day, also shares the same RM.
However, burst 200120A, detected a month later, shows a marginal increase in RM of $\sim 1$\,rad\,m$^{-2}$, which is still compatible at $2\sigma$ level.
Therefore, even though the varying contribution of the ionosphere only accounts for $\sim 0.1$\,rad\,m$^{-2}$, we consider the measurements in agreement and more observations are required to detect an eventual evolution of the RM with time.

Each of the bursts has been corrected for its RM and the resulting polarization profiles are reported in Figure \ref{fig:pol}.
Following \citet{everett2001}, we estimate that all the bursts in the sample have a linear polarization fraction larger than 90\% and a circular polarization fraction consistent with zero. However, we note that the fractional polarization measurements could be biased, especially for weak bursts.

The time-resolved polarization angle (PA), also shown in Figure \ref{fig:pol}, is approximately flat within each burst, with reduced $\chi^2$ values with respect to a straight line ranging between $0.7$ and $0.9$.
Due to the lack of an absolute polarization calibration, we did not attempt to compare PA values for different bursts, except for bursts 191219A and 191219B, which are separated by only 100\,ms. Their average PA values are in agreement, with a difference of the weighted PA curves of $0.5 \pm 2.5$\,deg.

It is interesting comparing the polarization properties of FRB 180916.J0158+65 with the other repeating FRB source with a measured RM, FRB 121102 \citep{michilli2018}.
The latter shows an extreme value of Faraday rotation corresponding to an $\text{RM} \sim 10^5$\,rad\,m$^{-2}$, orders of magnitude higher than the value measured for FRB 180916.J0158+65.
The RM of FRB 121102 varied by $\sim 10\%$ over 7 months, which is not observed for FRB 180916.J0158+65. 
The flat PA curve within each burst is an interested similarity between the two FRB sources, together with a linear polarization fraction close to 100\% and a circular polarization fraction of approximately 0\%.
The rest of the FRB population shows very diverse polarization properties, with varying RM values, polarization fractions and PA curves \citep{petroff2016}.

\subsection{Periodicity Search}
Although the source exhibits a modulation in activity at a 16.35-day period (or possibly a higher-frequency alias), an additional periodicity of the order of milliseconds to seconds might be present if the progenitor is a neutron star \citep{yang2020,lyutikov2020,ioka2020} and the bursts come from a narrow range of rotational phases. We use two different methods to search for a periodicity due to stellar rotation. First, we fit the largest common denominator to the differences in arrival times of bursts detected in individual observations, as is done for rotating radio transients \citep{mclaughlin2006}. We did not find any statistically significant periodicities in the two observations having multiple detections (see Table \ref{tab:observations}). We note that bursts 191219A and 191219B have a separation of $\sim$60 ms which is similar to the separation between sub-bursts in two bursts of this source, as noted by \citet{chime2019c}. However, periods close to 60 ms were not found to be statistically more significant than any other trial periods in the aforementioned analysis. Detection of eight or more bursts in a single observation might be necessary to determine an underlying periodicity, if it exists, with this method \citep{cui2017}.

For GBT observations in which there were single-pulse detections, we also searched for a rotational periodicity using PRESTO's Fourier domain acceleration search algorithm, \texttt{accelsearch}. We conducted this search on the same 327.68-$\mu$s resolution data as the initial single-pulse search, with the same \texttt{rfifind} mask and dedispersed to the same grid of DMs. Initially we searched all DMs using a \texttt{zmax} parameter of 200, implying an acceleration range of $\pm 20$\,m\,s$^{-2}$ for a 500\,Hz (2\,ms) signal. Next, we used the ``jerk-search'' functionality of \texttt{accelsearch} \citep{andersen2018} to search at the nominal DM with a \texttt{zmax} of 100 and a \texttt{wmax} of 500, corresponding to an acceleration range of $\pm 10$\,m\,s$^{-2}$ and a jerk range of $0.02$\,m\,s$^{-3}$ for a 500\,Hz signal. Finally, we ran a jerk search for the period of time in the MJD 58836 observation encompassing 10\,min around burst 191219B, with a \texttt{zmax} of 100 and a \texttt{wmax} of 500, corresponding to an acceleration range of $\pm 167$\,m\,s$^{-2}$ and a jerk range of $1.40$\,m\,s$^{-3}$ for a 500\,Hz signal. The best candidates from all three of these searches were folded using PRESTO's \texttt{prepfold} and inspected by eye. No convincing periodic astrophysical signals were found.

\begin{figure}
    \centering
    \includegraphics[scale=0.55]{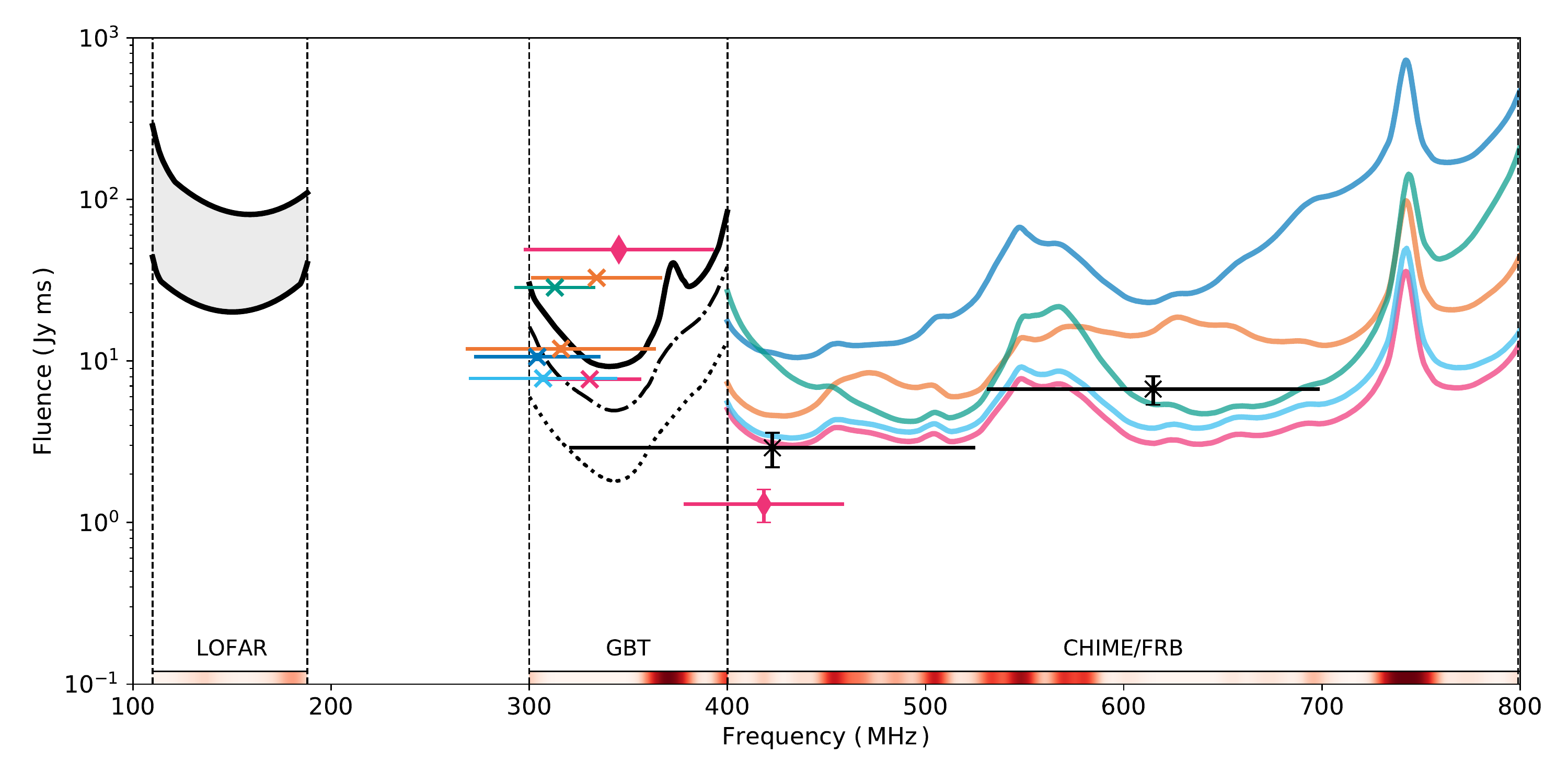}
    \caption{Burst fluences and completeness limits for LOFAR, GBT, and CHIME/FRB. Vertical dashed lines bracket the observing frequencies of each instrument. Markers with $\times$'s represent bursts detected with a single instrument, while the diamond markers correspond to the co-detection between GBT and CHIME/FRB. Markers are placed according to the peak frequency of their Gaussian fits, with horizontal lines in the frequency axis representing the FWTM of the fit. Marker colors and line colors are matched to show instantaneous sensitivity across instruments (approximated as constant for GBT and LOFAR, variable for CHIME/FRB depending on source position within synthesized beams). Solid lines show fluence completeness at the 90\% confidence interval towards bursts with Gaussian spectral profiles centered at each given frequency, with emission bandwidths ranging from 10--80 MHz (FWHM). The nonmonotonic nature of the CHIME/FRB completeness curves (i.e., crossing lines) are due to sidelobe sensitivity of the four synthesized beams relevant for detection. For GBT, we also show 50\% completeness and 5\% completeness with dash-dotted and dotted lines respectively, indicating typical and optimistic sensitivities. For LOFAR, we show completeness in the limit of no scattering (bottom line), and for the scattering timescale upper limit of 50 ms at 150 MHz (top line). The red shaded strips along the bottom of the plot show average RFI masking fractions of each instrument convolved with a 10 MHz (FWHM) Gaussian profile; masking fraction ranges from 0\% (white) to 100\% (dark red).}
    \label{fig:fluence}
\end{figure}

\subsection{Determination of System Sensitivity}\label{sec:fluence}
We determine fluence completeness for the CHIME/FRB system at the time of detection of the GBT bursts by expanding on the methods outlined in \cite{josephy2019}. Previously, detection scenarios were generated in a Monte Carlo simulation in order to capture sensitivity variations due to source position along transit, burst spectral shape, and detection epoch.
Relative sensitivities for these detection scenarios were translated to fluence thresholds using bandpass-calibrated observations of bursts detected from FRB \src by \citet{chime2020}. In this work, we do not sample the transit position when simulating detection scenarios; instead we examine the instantaneous sensitivity at the time of each GBT burst. This distinction is important for CHIME/FRB, since the 
sensitivity varies significantly depending on the source's position within the synthesized beams, as seen in Fig.~\ref{fig:fluence}. Furthermore, we compute fluence thresholds for Gaussian bursts of different central frequencies instead of 
computing band-averaged fluence thresholds. This choice is informed by the bands under consideration having frequency-dependent sensitivity and appreciable RFI contamination due to which the detectability of a given burst varies significantly as it is shifted throughout the band.
For burst bandwidths, we uniformly sample from 10--80 MHz (FWHM) to match the observed range. 

We extend this analysis to obtain the fluence completeness limit for the GBT observations by swapping in the appropriate bandpass.
The frequency response across the band 
is estimated to be the mean of the counts detected in each frequency channel over all time intervals not flagged for masking.
The completeness limit as a function of the central frequency of emission is shown in Figure \ref{fig:fluence}, while the band-averaged 90\%-completeness limit is estimated to be 22 Jy ms. In extending the analysis to LOFAR, we determine the frequency response using Equation \ref{eq:radiometer}; 
taking into account the exact number of Core stations used and correcting for the zenith-angle dependence as described by \citet{2016A&A...585A.128K}. Since no burst was detected in the LOFAR observations, we use the radiometer equation to construct a fiducial reference burst, which is parameterized as having the mean width of all bursts reported in this work (3.3 ms), a S/N of 7 (detection threshold of the search pipeline), a central frequency of 150 MHz, and a spectral bandwidth of 80 MHz (FWHM). 
With 18\% of channels having some RFI contamination, we take the effective bandwidth to be 64 MHz. 
Substituting these parameters and a band-averaged sky temperature, receiver temperature and gain in the radiometer equation in the limit of no pulse broadening, this reference scenario gives a fluence detection limit of 21 Jy ms. Considering the same range of Gaussian spectral profiles as above, the band-averaged 90\%-completeness limit is estimated to be 24 Jy ms. To account for suppressed detectability due to potential pulse broadening, we repeat this analysis using the upper limit on the scattering timescale of 50 ms at 150 MHz (see \S \ref{sec:scattering}). In this scenario, 
the band-averaged 90\%-completeness becomes 106 Jy ms, assuming a power-law index of $-$4 for the frequency dependence of scatter-broadening for different spectral profiles. The completeness threshold is significantly higher than that of the GBT due to the high Galactic foreground emission (a sky temperature of $\sim$700 K at 150 MHz) at the source position.

\section{Discussion}\label{sec:discussion}

The detection of emission in multiple bursts from FRB \src down to the bottom of the GBT band indicates that a cutoff or turnover in the FRB rest-frame spectrum can only exist below a frequency of $\sim$300 MHz. Since this source is the closest localized FRB \citep{marcote2020}, 
other FRBs could be detectable at frequencies lower than 300 MHz as their rest-frame spectra would be redshifted to even lower frequencies. We note that this conclusion would not apply to the so-far non-repeating FRBs if repeating and non-repeating sources do not share a common emission mechanism. 
A spectral turnover due to propagation effects in the circumburst environment \citep{ravi2019} or scattering combined with a flat spectrum for FRB emission 
(\citealt{chawla2017}) can still render FRBs undetectable at low frequencies. Therefore, detection of emission from this source at 300 MHz can potentially still be consistent with the non-detection of FRBs with several surveys in this frequency range \citep{deneva2016,chawla2017,rajwade2020limits} if a less dense circumburst environment (see \S \ref{sec:constraints}), low scattering timescale (see \S \ref{sec:scattering}) and the proximity of the source
conspire to make its emission
particularly detectable. 

\subsection{Scattering Times}\label{sec:scattering}
Assuming that the scattering properties of the intervening medium do not change between detections, we report a 95\% confidence upper limit of 1.7 ms on the scattering timescale of the source at 350 MHz, 
improving significantly on the constraint derived from CHIME/FRB observations, $<$ 0.9 ms at 600 MHz (which translates to $<$ 8 ms at 350 MHz, assuming a power-law frequency dependence of $-$4; \citealt{bhat2004}). An indirect measurement of scattering for this source, $\sim$2.7 $\mu$s at 1.7 GHz, was obtained by \citet{marcote2020} using the auto-correlation function estimated from a burst spectrum, 
assuming that the frequency-dependent intensity variations are due to scintillation. This translates to a scattering timescale of $\sim$1.5 ms at 350 MHz which is consistent with the GBT measurement. We used the methodology and code employed by \citet{masui2015} to estimate the auto-correlation function for the spectrum of the brightest burst in our sample (191219C) but find its magnitude small enough to suggest a non-astrophysical origin and be potentially explained by errors in modelling the burst spectrum or in correcting for the receiver bandpass.

Assuming a power-law index of $-$4, the upper limit on the scattering timescale
derived from GBT observations translates to 
50 ms at 150 MHz.  This suggests that pulse smearing due to scattering alone cannot explain the non-detection with LOFAR, for which 
observations were searched for bursts with widths up to 0.3 s (see \S \ref{sec:LOFAR}). Detection of the source at frequencies lower than 300 MHz is essential to obtain an accurate measurement of the scattering timescale. However, the measured upper limit is not inconsistent with that of known FRBs:  an FRB discovered by the UTMOST telescope exhibited a scattering timescale of 4.1$\pm$2.7 $\mu$s at 835 MHz (scales to 0.13 ms at 350 MHz; \citealt{farah2018}), suggesting that FRB \src is not atypical of the FRB population.  

\subsection{Phase Dependence of Source Activity}\label{sec:phase}
If the periodic modulation of source activity is due to orbital motion, then
any inhomogeneities in the source environment can cause some burst properties to be phase-dependent, as suggested by \citet{chime2020}. Following the approach of \citet{chime2020}, we compute the phases of all detections
and see no monotonic variation in DM, fluence, burst width, scattering timescale or emission frequency with phase or time in the sample of GBT bursts 
as well as in the overall sample of bursts from the source which includes detections from the CHIME/FRB system and the European Very-long-baseline-interferometry Network (EVN;  \citealt{marcote2020}). We caution that observations of the source with the GBT
did not cover a large fraction of the 5-day-long activity window. Additionally, we note that the aforementioned conclusion 
would not hold if the true period is not 16.35 days, as the estimated phases for all detections would be incorrect. Detections with the GBT cannot help in ruling out any of the possible higher-frequency aliases of the 16.35-day period, as all observations in which these detections were made were coincident with the transit of the source over the CHIME telescope.

\subsection{Frequency Dependence of Source Activity}\label{sec:spectralindex}
Six of the seven GBT bursts were not detected with the CHIME/FRB system although their GBT fluences were higher than or, in two cases, comparable to the corresponding 95\% confidence fluence threshold (see Figure \ref{fig:fluence}). The burst for which coincident emission was detected 
was the only GBT burst with emission detected at the top of the band (400 MHz), 
with a GBT-measured fluence of 48.9 $\pm$ 0.3 Jy ms and a fluence in the CHIME band of 1.3 $\pm$ 0.3 Jy ms. Coincident detection 
of only one burst out of a total of ten unique bursts detected by the two instruments during simultaneous observations suggests that emission is restricted to a narrow frequency range for individual bursts (fractional bandwidths of 13--30\% were measured for the GBT detections), with the overall source spectrum being highly variable as the center frequency of the individual bursts varies with time. 
This is consistent with the non-detection of coincident emission from FRB 121102 with the Arecibo telescope at 1.4 GHz and the Very Large Array (VLA) at 3 GHz for several bursts detected in simultaneous observations with the two instruments \citep{law2017,gourdji2019}.
Similarly, \citet{2019ApJ...876L..23H} used 1.1--1.7-GHz Arecibo observations to show that FRB 121102 bursts have fractional bandwidths ranging from 7--30\% at these frequencies. Longer observations of repeating FRB sources, simultaneously in several observing bands, are required to identify any potential temporal variations in the emission frequency of multiple narrow-band detections.

Based on a total of 3.9 hours of exposure to the source, we estimate the burst rate in the 300--400 MHz frequency range to be $\lambda_1 = 1.8^{+1.9}_{-1.1}$ bursts hr$^{-1}$ above a fluence threshold, $F_{\nu_1,\textrm{min}}$, of 22 Jy ms, with the uncertainties representing the 95\% confidence interval derived assuming Poisson statistics. Since all observations were within a phase window of 0.1 (phases ranging from 0.47 to 0.57; see Figure \ref{fig:exposure}), we compare the GBT-detection rate with the CHIME/FRB measurement in the same phase window, $\lambda_2 = 1.7^{+1.3}_{-0.8}$ bursts hr$^{-1}$ above a fluence threshold, $F_{\nu_2,\textrm{min}}$, of 5.2 Jy ms \citep{chime2020}. 
We scale the CHIME/FRB detection rate to that expected for a fluence threshold of $F_{\nu_1,\textrm{min}}$ using a power-law index, $\gamma = -2.3 \ \pm \ 0.4$, estimated for the differential energy distribution of the source by \citet{chime2020} and find that the expected detection rate is consistent with that measured for the GBT. Therefore, we cannot ascertain whether there is a variation in source activity with frequency using a direct comparison of the detection rates in the two bands.

Following \citet{houben2019}, we
characterize the frequency dependence of source activity using a 
statistical spectral index, $\alpha_\mathrm{s}$, which characterizes the power law relating the normalization (A) of the differential energy distribution (dN($\nu$)/dE = A($\nu$)E$^{\gamma}$) at different frequencies. The detection rates, $\lambda_1$ and $\lambda_2$, at two frequencies, $\nu_1$ and $\nu_2$, are then related as, 
\begin{equation}\label{equation:spectra}
\frac{\lambda_1}{\lambda_2} = \bigg(\frac{\nu_1}{\nu_2}\bigg)^{-\alpha_\mathrm{s} \gamma} \bigg(\frac{F_{\nu_1,\textrm{min}}} {F_{\nu_2,\textrm{min}}}\bigg)^{\gamma+1}.
\end{equation} 
We estimate $\alpha_\mathrm{s} = -1.6^{+1.0}_{-0.6}$, 
with the reported 95\% confidence level uncertainties 
determined by the range obtained when solving for this parameter in 10,000 simulations of sets of $\lambda_1$, $\lambda_2$ and $\gamma$, where each is sampled from its Gaussian distribution of mean and range given above.
This measurement is consistent within reported uncertainties with the statistical spectral index estimated for FRB 121102 using observations
in the frequency range of 1.2--3.5 GHz by \citet{houben2019}. 
We note that this measurement is robust only if the true period of the source is 16.35 days since a higher-frequency alias would imply that all observations are at phases different than those currently estimated. Moreover, the conclusion also relies on the burst rate being constant in each period of source activity since the GBT detection rate is estimated from observations over three periods as compared to the $\sim$20 periods used for the CHIME/FRB measurement. 
If the aforementioned assumptions hold, then the measurement implies decreasing burst activity with increasing frequency for the source and could explain its non-detection at 1.4 GHz (above a fluence threshold of 0.17 Jy ms) with the Effelsberg telescope in 17.6 hours of exposure during a predicted epoch of activity \citep{chime2020}. This conclusion would also be consistent with the spectral index of $\alpha=-1.5\pm0.2$ estimated for the mean spectrum of 23 FRBs detected by the Australian SKA Pathfinder at 1320 MHz \citep{macquart2019} with the caveat that the mean spectrum is not redshift-corrected while the statistical spectral index that we evaluate characterizes the rest-frame spectrum of the source.

We did not detect coincident emission with LOFAR at the time of detection of six GBT bursts. Assuming the emission is broadband and that the burst fluence scales with frequency as $F_\nu \propto \nu^\alpha$, we can use the brightest burst in our sample, 191219C, to place a constraint on the spectral index, $\alpha$, of the instantaneous emission from the source. We obtain a lower limit of $\alpha > -1.0$, which is the spectral index for which the 95\% confidence lower bound on the GBT-measured fluence of the burst (see Table \ref{tab:bursts}) would imply a fluence at 150 MHz equal to the 90\% completeness fluence threshold of the LOFAR observations (106 Jy ms; see \S\ref{sec:fluence}). However, it could be that coincident emission was suppressed in the LOFAR band not due to the burst having a flat spectrum but rather due to its band-limited nature. In that scenario, non-simultaneous bursts could still have been detected in the LOFAR band during the 1.3 hours of exposure to the source. The non-detection of such bursts could be explained by flattening of the statistical spectrum at low frequencies, as suggested by \citet{houben2019} for FRB 121102. To investigate the possibility of spectral flattening, we derive a 95\% confidence upper limit on the burst rate in the LOFAR band of 2.2 bursts hour$^{-1}$ above a fluence threshold of 106 Jy ms.  We then obtain a 95\% confidence lower limit on the statistical spectral index between the GBT and LOFAR bands, $\alpha_\mathrm{s} > -1.4$. Since this measurement of the statistical spectral index is consistent with that derived by comparing detection rates in the GBT and CHIME bands, we cannot confirm or rule out a flattening of the spectrum at low frequencies.

\subsection{Constraints on Proposed Models}\label{sec:constraints}
The detection of FRB 180916.J0158+65 down to 300 MHz shows that its source's ambient environment is optically thin to free-free absorption at 300 MHz. For an ionized nebula of size L$_{\text{pc}}$, and  DM $<$ 70 pc cm$^{-3}$ \citep{marcote2020}, this implies the optical depth due to free-free absorption \citep{condon2016essential},
\begin{equation}
\tau_{\text{ff}} = 1.6 \times 10^{-3} \times \Big( \frac{\textrm{T}}{10^{4} \text{K}} \Big)^{-1.35} \times \Big( \frac{\nu}{1 \text{GHz}} \Big)^{-2.1} \times \frac{1}{\text{f}_{\mathrm{eff}} \mathrm{L}_{\text{pc}}} \times \Big( \frac{\mathrm{DM}}{70 \text{ pc cm}^{-3}}\Big)^{2} \ll 1,
\label{Eq:ff}
\end{equation} where  $\text{f}_{\mathrm{eff}}$ is a factor that accounts for the volume-filling factor and the electron density fluctuation in the circumburst medium.  
If such a nebula  circumscribes the FRB 180916.J0158+65 source, 
 our detection implies L $\gg$ 0.02 pc (T/$10^4$~K)$^{-1.35}$
which rules out a surrounding dense and compact ionized nebula like
a hyper-compact H\textsc{ii} region \citep{churchwell2002ultra} or a young supernova remnant (age $<$ 50 yr; \citealt{piro2016impact}). This result is consistent with observations of the source by \citet{marcote2020} who conclude, based on the measured RM and non-detection of an associated persistent radio source, that FRB 180916.J0158+65 has a much less extreme circumburst environment compared to that of the other localized repeater, FRB 121102.
From Equation \ref{Eq:ff}, free-free absorption at frequencies 
well below 300 MHz seems
unlikely to inhibit detectability, so
follow-up observations in this regime are
warranted in order to constrain the intrinsic FRB spectrum turn-over frequency. 

Induced Compton scattering (ICS) can also suppress the observed low-frequency emission from FRBs specifically in models that invoke magnetar giant flares \citep{lyubarsky2014model,beloborodov2017flaring,metzger2019fast,margalit2019constraints,levin2020precessing}.
In the synchrotron maser model
\citep{metzger2019fast,margalit2019constraints},
FRBs are produced when the decelerating ultra-relativistic shock waves from a flaring hyperactive magnetar interact with the sub-relativistic circumstellar medium (CSM). As the shock propagates into the CSM, ICS attenuates the FRB emission. However, the negative spectral indexes of some of the bursts we have observed for FRB 180916.J0158+65 argue that the optical depth due to 
ICS, $\tau_\mathrm{ICS} < 1$ at 300 MHz.
The constraint that $\tau_{\mathrm{ICS}} <$ 1 implies a density upstream from the shock \citep{pawan2020ics},
\begin{equation}
    \rho \lesssim 8 \times 10^{-3} \mathrm{cm}^{-3} \Big(\mathrm{r}_{\mathrm{dec}}/{10^{12}} \mathrm{cm}\Big)^{2} 
    \Big(\mathrm{L}{/10^{43} \mathrm{erg \; s^{-1}} \Big)^{-1} 
    \Big(\delta t/1 \mathrm{ms}\Big)^{-1} \Big(\mathrm{\Gamma_{\mathrm{CSM}}}}\Big)^{-2},
\end{equation}
at the deceleration radius r$_{\mathrm{dec}}$, the distance that a relativistic 
shock
travels before half of its energy is dissipated to its ambient medium,  $\sim$ 10$^{12}$ cm \citep{metzger2019fast}. Here, L is the isotropic equivalent luminosity of the FRB, $\delta$t is the burst temporal width, and $\Gamma_{\mathrm{CSM}}$ is the Lorentz factor of the shocked CSM gas. 
However, the \citet{metzger2019fast} model requires an upstream density of $\sim$ 10$^{2}$ cm$^{-3}$ for
the synchrotron maser to be effective; for our optical depth constraint, this requires
r$_{\mathrm{dec}}$ $>$ 10$^{14}$~cm for the above
parameters.
\cite{pawan2020ics} have argued that it is 
difficult
to produce FRBs at distances $> 10^{14}$ cm because of the 
requirement of the source luminosity 
to be $> 10^{46}$ erg s$^{-1}$ for each repeat burst --- difficult to obtain even from a hyperactive magnetar. 
Thus, the detection of FRB 180916.J0158+65 bursts down to 300 MHz 
challenges some of
the 
assumptions of the maser synchrotron model.

\section{Conclusions}\label{sec:conclusion}
We have reported on the detection of emission from the repeating FRB \src at frequencies down to 300 MHz. 
This result implies that any cutoff in the rest-frame FRB spectrum exists at a frequency below 300 MHz, and thus bodes well for future blind surveys for FRBs as well as follow-up observations of repeating FRBs at low radio frequencies. Using burst rates in the GBT band (300--400 MHz) and the CHIME band (400--800 MHz), we compute a statistical spectral index, $\alpha_\textrm{s}$, to characterize the frequency dependence of source activity. We find $\alpha_\textrm{s} = -1.6^{+1.0}_{-0.6}$, which is consistent with observations of FRB 121102 by \citet{houben2019} and motivates follow-up observations of other repeating FRBs to determine whether this trend is common to all sources. We do not see any frequency dependence to the polarization fraction in the overall sample of bursts from this source, with the four brightest bursts in the GBT sample being nearly 100\% linearly polarized and having a circular polarization fraction consistent with zero. 

Our observation strategy of simultaneous coverage with GBT, the CHIME/FRB system and LOFAR allowed for a study of emission from this source over a large range of frequencies. Most bursts detected during these observations had low emission bandwidths (being observable with either the GBT or with CHIME/FRB) with the notable exception of GBT-detected burst 191219C which had a potentially associated sub-burst in the CHIME band. We detected no bursts in the frequency range 110--190 MHz with LOFAR despite performing an extensive search for bursts with narrow emission bandwidths and slow frequency drifts by visually inspecting LOFAR data in a $\pm$10-s interval around the 
arrival times of the GBT bursts. 
Our constraint on the scattering timescale for the source, $<$ 1.7 ms at 350 MHz, rules out the phenomenon of scattering alone as a cause of non-detection of the source with LOFAR. 
However, we cannot yet say whether the reduced system sensitivity owing to the high sky temperature or a spectral turnover at these frequencies was the cause of the non-detection. 
More observations of the source at lower frequencies might help determine whether a spectral turnover exists. 

In the late stages of writing this paper, we became aware of follow-up observations of the source at 328 MHz with the Sardinia radio telescope by \citet{pilia2020}. Detection of three bursts from the source in these observations is consistent with our conclusions.

\acknowledgments
We thank the Dominion Radio Astrophysical Observatory, operated by the National Research Council Canada, for gracious hospitality and useful expertise. CHIME is funded by a grant from the CFI Leading Edge Fund (2012) 
(Project 31170)  and by contributions from the provinces British 
Columbia, Quebec and Ontario. The CHIME/FRB Project is funded by a grant from the Canada Foundation for Innovation 2015 Innovation Fund (Project 33213), as well as by the Provinces of British Columbia and Qu\'ebec, and by the Dunlap Institute for Astronomy and Astrophysics at the University of Toronto. Additional support was provided by the Canadian Institute for Advanced Research (CIFAR), McGill University and the McGill Space Institute via the Trottier Family Foundation, and the University of British Columbia. Research at Perimeter Institute is supported by the Government of Canada through Industry Canada and by the Province of Ontario through the Ministry of Research \& Innovation. The National
Radio Astronomy Observatory is a facility of the National Science Foundation (NSF) operated under cooperative agreement by Associated Universities, Inc. The Green Bank Observatory is a facility of the National Science Foundation (NSF) operated under cooperative agreement by Associated Universities, Inc. We thank Compute Canada, the McGill Center for High Performance Computing, and Calcul Qu\'ebec for provision and maintenance of the Beluga supercomputer and related resources.
This paper is based (in part) on data obtained from facilities of the International LOFAR Telescope (ILT) under project code DDT12\_001.
LOFAR~\citep{2013A&A...556A...2V} is the Low Frequency Array designed and constructed by ASTRON. It has observing, data processing, and data storage facilities in several countries, that are owned by various parties (each with their own funding sources), and that are collectively operated by the ILT foundation under a joint scientific policy. The ILT resources have benefited from the following recent major funding sources: CNRS-INSU, Observatoire de Paris and Universit\'e d'Orl\'eans, France; BMBF, MIWF-NRW, MPG, Germany; Science Foundation Ireland (SFI), Department of Business, Enterprise and Innovation (DBEI), Ireland; NWO, The Netherlands; The Science and Technology Facilities Council, UK; Ministry of Science and Higher Education, Poland. 
The research leading to these results has received funding from the European Research Council under the European Union's Seventh Framework Programme (FP7/2007-2013) / ERC grant agreement nr. 337062 (DRAGNET; PI: Hessels).

P.C. and M.B. are supported by an FRQNT Doctoral Research Award. V.M.K. holds the Lorne Trottier Chair in Astrophysics \& Cosmology and a Canada Research Chair and receives support from an NSERC Discovery Grant and Herzberg Award, from an R. Howard Webster Foundation Fellowship from the Canadian Institute for Advanced Research (CIFAR), and from the FRQNT Centre de Recherche en Astrophysique du Quebec. D.M. is a Banting Fellow. Z.P. is supported by a Schulich Graduate Fellowship. The Dunlap Institute is funded through an endowment established by the David Dunlap family and the University of Toronto. FRB research at UBC is supported by an NSERC Discovery Grant and by the Canadian Institute for Advanced Research. M.D. is supported by a Killam Fellowship and receives support from an NSERC Discovery Grant, the Canadian Institute for Advanced Research (CIFAR), and from the FRQNT Centre de Recherche en Astrophysique du Quebec. B.M.G. acknowledges the support of the Natural Sciences and Engineering Research Council of Canada (NSERC) through grant RGPIN-2015-05948, and of the Canada Research Chairs program. J.W.T.H. acknowledges funding from an NWO Vici fellowship. P.S. is a Dunlap Fellow and an NSERC Postdoctoral Fellow. 

\vspace{5mm}
\facilities{GBT, CHIME, LOFAR}

\software{PRESTO \citep{ransom2001}, \texttt{cdmt} \citep{2017A&C....18...40B}, \texttt{DM\_phase} \citep{seymour2019}, \texttt{PSRCHIVE} library \citep{hotan2004}, \texttt{RMextract} \citep{2018ascl.soft06024M}}

\bibliography{sample63}{}
\bibliographystyle{aasjournal}


\end{document}